\documentclass[a4paper,onecolumn]{IEEEtran}

\usepackage[utf8]{inputenc} 
\usepackage[T1]{fontenc}
\usepackage{url}
\usepackage{ifthen}
\usepackage{cite}
\usepackage[cmex10]{amsmath} 
\usepackage{amssymb}
\usepackage{amstext}

\usepackage{graphics}
\usepackage{tikz}
\usetikzlibrary{positioning,chains,fit,shapes,calc,arrows}
\usepackage{verbatim,booktabs}
\usepackage{notation}
\usepackage{dsfont}

\newtheorem{theorem}{Theorem}
\newtheorem{corollary}{Corollary}
\newtheorem{lemma}{Lemma}

\IEEEoverridecommandlockouts

\begin{document}
\title{Minimum Probability of Error of \\List $M$-ary Hypothesis Testing} 


\author{Ehsan Asadi Kangarshahi and Albert Guill\'en i F\`abregas
\thanks{E. Asadi Kangarshahi is with the Department of Engineering, 
University
of Cambridge, Cambridge CB2 1PZ, U.K. (e-mail: ea9972@gmail.com).

A.~Guill\'en i F\`abregas is with the Department of Engineering, 
University
of Cambridge, Cambridge CB2 1PZ, U.K. and also with the Department of Information and 
Communication
Technologies, Universitat Pompeu Fabra, Barcelona 08018, Spain (e-mail: guillen@ieee.org).

This work was supported in part by the European Research Council under 
Grant 725411.
}
}


\maketitle

\begin{abstract}
We study a variation of Bayesian $M$-ary hypothesis testing in which the test outputs a list of $L$ candidates out of the $M$ possible upon processing the observation. We study the minimum error probability of list hypothesis testing, where an error is defined as the event where the true hypothesis is not in the list output by the test. We derive two exact expressions of the minimum probability or error. The first is expressed as the error probability of a certain non-Bayesian binary hypothesis test, and is reminiscent of the meta-converse bound. The second, is expressed as the tail probability of the likelihood ratio between the two distributions involved in the aforementioned non-Bayesian binary hypothesis test.
\end{abstract}


\section{Introduction}


Statistical hypothesis testing is the problem of deciding one of $M$ possible statistical hypotheses after processing some observation data modeled by a random variable. Hypothesis testing is one of the main problems in statistics and inference and finds applications in areas such as social, biological, medical, computer sciences, signal processing and information theory. Depending on the subject area and underlying assumptions, it can be referred to as model selection, classification, discrimination or detection.
Hypothesis testing problems are typically classified as binary or non-binary, depending on the number of hypotheses, and Bayesian or non-Bayesian, depending on whether or not priors on the hypotheses are known.

The minimum average probability of error of Bayesian binary hypothesis testing is attained by the likelihood ratio test. Similarly, the minimum average error probability of Bayesian $M$-ary hypothesis testing is attained by the maximum a posteriori (MAP) test \cite{lehmann2005testing}. For non-Bayesian binary hypothesis testing, Neyman and Pearson formulated the optimal tradeoff between the pairwise error probabilities and showed that the likelihood ratio test attains the optimum tradeoff \cite{neyman1933ix}. 

We study a variation of Bayesian $M$-ary hypothesis testing. Specifically, we allow the test to output a list with $L$ candidate hypotheses. This setting is helpful when the number of hypotheses is very large and, for complexity reasons, one might wish to implement staggered or iterative testing. At each stage, a bank of tests of smaller dimension is run, but a candidate list is output instead of a single candidate, in order to facilitate information exchange at the next stage or iteration. List hypothesis testing is also implicitly employed in approximate recovery problems related to statistical estimation where a reduction to multiple hypothesis testing is performed (see e.g. \cite[Sec. 16.2.2.]{scarlett2021introductory}). In reliable data transmission or storage, list decoding is employed in order to improve the performance of error-correcting codes \cite{elias1957list}. In communications, list detection is employed in large linear multiple-input multiple-output systems iteratively exchanging information with iterative decoders of error correcting codes (see e.g. \cite{studer2010soft}). 

From a theoretical perspective, it is important to understand what is the minimum error probability in order to establish a performance benchmark for practical tests. In this paper, we study the minimum probability of error of list hypothesis testing. We provide two new families of bounds to the minimum probability of error. The first one, bounds the minimum probability of error by that of a suitably optimized non-Bayesian hypothesis test and is reminiscent of the meta-converse bound in \cite{polyanskiy2010channel}. Instead, the second family bounds the minimum probability of error by the tail probability of the likelihood ratio, or the information spectrum  \cite{han2003information}. When these bounds are optimized over an auxiliary output distribution, inspired by the work in \cite{vazquez2016bayesian}, we show that the bounds are actually tight and provide two different expressions of the minimum probability of error
We show that the solution of the optimization of the second bound is unique and provide an expression for the optimal auxiliary distribution. In turn, the identities not only help in better understanding the minimum probability of error, but also help assessing the tightness of the bounds.

This paper is structured as follows. Section \ref{sec:binary} introduces the relevant notation for binary hypothesis testing. Section \ref{sec:list} describes the list hypothesis testing problem and derives the minimum probability of error. Section \ref{sec:meta} proves the first identity for the minimum probability of error and connects it with non-Bayesian binary hypothesis testing. Section \ref{sec:vh} proves the second identity for the minimum probability of error and connects it information spectrum. In proving this result, it is shown that the optimal auxiliary distribution is unique. Proofs of auxiliary results can be found the Appendix.


\section{Binary Hypothesis Testing}
\label{sec:binary}
Let $Y$ be a random variable taking values on set $\Yc$. We consider two hypotheses $H$, $0$ and $1$, which correspond to $Y$ being distributed according to two distributions, $P$ or $Q$, respectively. A binary hypothesis test is a probabilistic  mapping $\Yc \to [0,1]$ that upon observing a certain $y$, decides which of the hypothesis represents the observation. We let $\hat H$ be the random variable associated with the output of test and $T$, the test mapping, the conditional distribution $P_{\hat H|Y}$.

The performance of binary hypothesis test is characterized by the type $0$ and type $1$ errors defined as follows:
\begin{align}
	&\epsilon_0(T,P) = \PP \big[\hat{H} = 1|H=0\big] = \sum_{y} P(y) T(1|y)\\
	&\epsilon_1(T,Q) = \QQ \big[\hat{H} = 0|H=1\big] = \sum_{y} Q(y)T(0|y).
\end{align}

In the Bayesian setting, given a prior probability $P_H(0), P_H(1)$, the smallest average probability of error is given by
\begin{align}
\bar \epsilon = \min_{T} \Big\{P_H(0)\cdot\epsilon_0(T,P) + P_H(1)\cdot\epsilon_1(T,P) \Big\}
\end{align}
and $T$ is known to be the likelihood ratio test \cite{lehmann2005testing}; the likelihood ratio $\frac{P(y)}{Q(y)}$ is checked against the ratio of the priors.

In the non-Bayesian setting, no knowledge about the prior probabilities $P_H(h_i)$, $i=0,1$ is assumed. The trade-off between the pairwise error probabilities $\epsilon_0(T,P)$ and $\epsilon_1(T,Q)$ is characterized by the function $\alpha_{\beta}(P,Q)$ defined as follows
\begin{align} \label{Neyman-Pearson}
	\alpha_{\beta}(P,Q) = \min_{T:\epsilon_1(T,Q) \leq \beta} \epsilon_0(T,P).
\end{align}
Similarly, one can define the alternative tradeoff $\beta_{\alpha}(P,Q)$ as
\begin{align} \label{vsmfuwffufwug}
	\beta_{\alpha}(P,Q) = \min_{T:\epsilon_0(T,P) \leq \alpha} \epsilon_1(T,Q).
\end{align}

It is well known that a minimizing test for \eqref{Neyman-Pearson} is the likelihood-ratio threshold test \cite{neyman1933ix}. It is known that every optimal test is a threshold test where the likelihood ratio between the two distributions is compared to a threshold  $\lambda_\text{NP}\in \RR$ such that optimal test can be expressed by the following equation
\begin{align}
	T_\text{NP}(1|y) = \begin{cases}
	1 \ &\frac{P(y)}{Q(y)} > \lambda_\text{NP} \\
	0 \ &\frac{P(y)}{Q(y)} < \lambda_\text{NP} \\
	\delta_y \ &\frac{P(y)}{Q(y)} = \lambda_\text{NP}
	\end{cases}
\end{align}
where in order to solve \eqref{Neyman-Pearson}, $\delta_y$ and $\lambda_{\text{NP}}$ are chosen such that $\epsilon_1(T,Q) = \beta$. The minimizing test is not unique in general since all values of $\delta_y$ and $\lambda_{\text{NP}}$ with the property $\epsilon_1(T,Q) = \beta$ yield an optimal test.

\section{List Hypothesis Testing}
\label{sec:list}

Consider now a Bayesian $M$-ary hypothesis testing problem, with two random variables $Y,X$ jointly distributed according to $P_{XY}$, such that $Y,X$ take values on $\Yc,\Xc$, respectively with $|\Xc| = M$. The observation alphabet $\Yc$ is a general alphabet that encompasses the Cartesian product of $n$-observations and many other standard settings. 
Upon observing $y \in \Yc$ we wish to decide what $X$ was. Standard $M$-ary hypothesis tests output a single candidate hypothesis $\hat X\in\{1,\dotsc,M\}$. Instead, we consider list hypothesis testing. A list hypothesis test with list size $L$ is a possibly random mapping $P_{\hat\Xm|Y}$, where $\hat \Xm = (\hat X_1,\dotsc,\hat X_L)\in\Xc^L$ denotes the random vector containing a list of candidates $\{\hat X_1,\dotsc,\hat X_L\}$. For simplicity of the presentation, we assume that all candidates in the list are distinct; this does not have an effect on the structure of the test that minimizes the probability of error.
We say that the true hypothesis has been successfully estimated if the true $X$ is one of the entries of the list vector $\hat \Xm =(\hat X_1,\dotsc,\hat X_L)$, i.e., if $X\in\{\hat X_1,\dotsc,\hat X_L\}$. The problem is obviously of interest when $L \ll M$.

Since the joint distribution $P_{XY}$ defines
a prior distribution $P_X$ over the alternatives, the problem is naturally cast within the Bayesian framework. The average probability of error of a given list hypothesis test $P_{\hat\Xm|Y}$, defined as $\bar \epsilon(P_{\hat\Xm|Y})$, is written as 
\begin{align}
\bar \epsilon(P_{\hat\Xm|Y}) &\eqdef \PP\big[X \notin \{\hat X_1,\dotsc,\hat X_L\}\big] \label{eq:prob1}\\
&=\PP\big[\{\hat X_1\neq X\} \cap  \cdots \cap \{\hat X_L \neq X\} \big]\label{eq:prob2}\\
&=1-\PP\big[\{\hat X_1= X\} \cup  \cdots \cup \{\hat X_L = X\} \big]\label{eq:prob3}\\
&=1-\EE\big[\openone\big\{\{\hat X_1= X\} \cup  \cdots \cup \{\hat X_L = X\} \big\}\big]\label{eq:prob4}\\
&= 1 -  \sum_{\substack{x\in\Xc,y\in\Yc\\\hat x_1,\dotsc,\hat x_L \in \Xc}} P_{XY}(x,y) P_{\hat \Xm |Y}(\hat x_1,\dotsc,\hat x_L|y)\openone\big\{\{\hat x_1= x\} \cup  \cdots \cup \{\hat x_L = x\} \big\}\label{expression}\\
&=1- \sum_{\substack{y\in\Yc\\\hat x_1,\dotsc,\hat x_L \in \Xc}} P_{\hat \Xm |Y}(\hat x_1,\dotsc,\hat x_L|y) \PP\big[X \in \{\hat x_1,\dotsc,\hat x_L\}, Y=y\big]
\end{align}	
where the probability and expectation in \eqref{eq:prob1}-\eqref{eq:prob4} is computed with respect to the joint distribution between the true hypothesis $X$, the observation $Y$ and the list $\hat\Xm$, and where
\begin{align}
\PP\big[X \in \{\hat x_1,\dotsc,\hat x_L\}, Y=y\big] &= \sum_{x \in \Xc} P_{XY}(x,y)\openone\big\{\{\hat x_1= x\} \cup  \cdots \cup \{\hat x_L = x\} \big\}\\
& = P_{XY}(\hat x_1,y) + \cdots + P_{XY}(\hat x_L,y).
\label{eq:plist}
\end{align}
Equation \eqref{eq:plist} holds since all elements on the list are assumed to be distinct, and thus, the events $\{\hat X_\ell = X\}$ for $\ell=1,\dotsc,L$, are disjoint.

%

Further define 
\begin{align} \label{Listdist}
P_{\Xm Y}(x_1,\dotsc,x_L,y) \eqdef \frac{1}{ \binom{M-1}{L-1}}\big(P_{XY}(x_1,y) +  \cdots + P_{XY}(x_L,y) \big)  
\end{align}
where $\binom{a}{b} = \frac{a!}{b!(a-b)!}$. Observe that \eqref{Listdist} is, by assumption, defined only for distinct $x_1,\dotsc,x_L\in\Xc$. 
In order to show that the above definition induces a probability distribution on $\Xc^L$, we write
\begin{align}
	\sum_{ x_1,\dotsc, x_L,y} P_{ \Xm Y}( x_1,\dotsc, x_L,y) &= \sum_{ x_1,\dotsc, x_L,y}\frac{1}{ \binom{M-1}{L-1}}\big(P_{XY}( x_1,y)  + \cdots + P_{XY}( x_L,y) \big) \label{eq:pxhatsum}\\
	& = \frac{1}{ \binom{M-1}{L-1} } \sum_{ x,y} \binom{M-1}{L-1} P_{XY}( x,y)  \label{eq:pxhatsum2}\\
	& = 1,
\end{align}
where \eqref{eq:pxhatsum} follows from the definition of $P_{ \Xm Y}( x_1,\dotsc, x_L,y)$ in \eqref{Listdist}, \eqref{eq:pxhatsum2} follows from the fact that for any given $x\in\Xc$ in the sum \eqref{eq:pxhatsum2}, there are $\binom{M-1}{L-1}$ possible list configurations. 

We now turn to the minimum probability of error over all tests, defined as 
\begin{align}
\bar \epsilon = \min_{P_{\hat \Xm|Y}} \bar{\epsilon}(P_{\hat \Xm|Y}).
\end{align}
The following result finds a test that achieves the minimal probability of error.

\begin{lemma}\label{lemma:minpe}
An optimal test achieving the minimal probability of error $\bar \epsilon$ chooses distinct $(\hat x_1,\dotsc,\hat x_L)\in\Xc^L$ such that $P_{  \Xm Y}(\hat x_1,\dotsc, \hat x_L,y)$ is maximized, yielding
\begin{align}
\bar{\epsilon} = 1-\binom{M-1}{L-1}\sum_{y \in \Yc} \max_{(\hat x_1,\dotsc,\hat x_L) \in \Xc^L} P_{ \Xm Y}(\hat x_1,\dotsc,\hat x_L,y).
\end{align}
\end{lemma}
\begin{IEEEproof}
Any test that maximizes $P_{ \Xm Y}(\hat x_1,\dotsc,\hat x_L,y)$ will maximize the probability of success and thus minimize the probability of error. Thus, we set
\beq
P_{\hat \Xm|Y} (\hat x_1,\dotsc,\hat x_L|y) = \begin{cases}
\frac{1}{|\Sc(y)|} & (\hat x_1,\dotsc,\hat x_L)\in\Sc(y)\\
0 &\text{otherwise}
\end{cases}
\eeq
where
\beq
\Sc(y) = \Big\{( x_1,\dotsc, x_L)\in\Xc^L \ : \  P_{ \Xm Y} ( x_1,\dotsc, x_L,y) = \max_{(\hat x_1,\dotsc,\hat x_L)\in\Xc^L}P_{ \Xm Y}(\hat x_1,\dotsc,\hat x_L,y) \Big\}
\label{eq:setS}
\eeq
is the set of list vectors that maximize $P_{ \Xm Y}$; there might be more than one maximizing list. With this particular choice, we obtain
\begin{align}
\bar\epsilon(P_{\hat\Xm|Y})  
& = 1-\sum_{\substack{y\in\Yc\\\hat x_1,\dotsc,\hat x_L \in \Xc}} P_{\hat \Xm |Y}(\hat x_1,\dotsc,\hat x_L|y) \big(P_{XY}(\hat x_1,y) + \cdots + P_{XY}(\hat x_L,y) \big)\\
& = 1-	\sum_{y\in\Yc} \sum_{\hat x_1,\dotsc,\hat x_L \in \Sc(y)} \frac{1}{|\Sc(y)|} \max_{(\hat x_1,\dotsc,\hat x_L) \in \Xc^L} \big(P_{XY}(\hat x_1,y) + \cdots + P_{XY}(\hat x_L,y) \big)\\
& = 1-	\sum_{y\in\Yc} \max_{(\hat x_1,\dotsc,\hat x_L) \in \Xc^L} \big(P_{XY}(\hat x_1,y) + \cdots + P_{XY}(\hat x_L,y) \big)\sum_{\hat x_1,\dotsc,\hat x_L \in \Sc(y)} \frac{1}{|\Sc(y)|} \\
&= 1-\sum_{y\in\Yc} \max_{(\hat x_1,\dotsc,\hat x_L) \in \Xc^L} \big(P_{XY}(\hat x_1,y) + \cdots + P_{XY}(\hat x_L,y) \big).
\end{align}
The final result is obtained from definition \eqref{Listdist}. Finally, observe that in order for the optimal test to maximze 
$\PP\big[X \in \{\hat x_1,\dotsc,\hat x_L\}, Y=y\big]$ 
it is needed that $\hat x_\ell$ for $\ell=1,\dotsc,L$ are distinct, since otherwise there would be fewer than $L$ summands in \eqref{eq:plist}.
\end{IEEEproof}

\section{Meta-Converse}
\label{sec:meta}

In reference \cite{polyanskiy2010channel} Polyanskiy, Poor and Verd\'u introduced a lower bound to the minimum probability of error of conventional $M$-ary hypothesis testing. The bound, termed meta-convertse, is expressed as the error probability of a non-Bayesian binary hypothesis test as
\beq
\bar \epsilon \geq \alpha_{\frac 1M}(P_{XY},Q_X\times Q_Y)
\eeq
where $Q_X(x)=\frac 1M$ for every $x\in\Xc$ and $Q_Y$ is an arbitrary auxiliary output distribution. It was shown in \cite{vazquez2016bayesian} that optimizing over $Q_Y$ results in the bound being tight thus providing the exact minimum probability of error. In this section, we show a similar family of bounds for list hypothesis testing and provide an identity that connects the minimum error probability of list hypothesis testing and the proposed bound by means of an optimization over the auxiliary distribution.

First, define an auxiliary probability distribution over the list vector
\begin{align}
Q_{\Xm}(x_1,\dotsc, x_L)  \eqdef
\begin{cases} \frac{1}{\binom{M}{L} } \  \ & \text{for distinct }  x_1,\dotsc,  x_L \in \Xc\\
 0 \ &\text{otherwise}
\end{cases} 
\label{eq:qdist}
\end{align}
where  $\Xm$ is a random vector defined on $\Xc^L$.

The following theorem states the main result of this paper for list hypothesis testing.

\begin{theorem}\label{thm:exact}
The minimum probability of error $\bar \epsilon$ of Bayesian $M$-ary list hypothesis testing with list size $L$ can be bounded as
\begin{align} 
\frac{1}{\binom{M-1}{L-1}}(1 - \bar \epsilon) &\leq 1 - \alpha_{\frac{1}{\binom{M}{L} }}(P_{ \Xm Y} , Q_{ \Xm} \times Q_Y)\label{eq:mcbound}
\end{align}
where $P_{ \Xm Y}$ and $Q_{ \Xm}$ are defined in \eqref{Listdist} and \eqref{eq:qdist}, respectively and $Q_Y$ is an arbitrary distribution over the observation alphabet $\Yc$. In addition,
\begin{align} 
\frac{1}{\binom{M-1}{L-1}}(1 - \bar \epsilon) &= 1 - \max_{Q_Y} \alpha_{\frac{1}{\binom{M}{L} }}(P_{ \Xm Y} , Q_{ \Xm} \times Q_Y)\label{eq:mc}
\end{align}
where the following distribution is a maximizer for expression \eqref{eq:mc}
\begin{align}
Q^*_{Y}(y) \triangleq \frac{1}{\mu}\max_{( x_1,\dotsc, x_L) \in \Xc^L} P_{ \Xm Y}( x_1,\dotsc, x_L,y)
\label{eq:qstar}
\end{align}
with
\beq
\mu = \sum_{y' \in \Yc} \max_{( x_1,\dotsc, x_L) \in \Xc^L} P_{ \Xm Y}( x_1,\dotsc, x_L,y')
\eeq
being a normalization constant.
\end{theorem}

\begin{IEEEproof}
By assuming that $P_{\hat \Xm|Y}$ is a probabilistic function for estimating $X$ with a list $\hat\Xm\in\Xc^L$, we have for all $y \in \Yc$ 
\begin{align} \label{anuetbge}
\sum_{\hat x_1,\dotsc,\hat x_L \in \Xc} P_{\hat \Xm|Y}(\hat x_1,\dotsc,\hat x_L|y) = 1.
\end{align}


%

We proceed by defining a binary hypothesis test $T$ between two distributions on $\Xc^L\times \Yc$ such that
\begin{align}
&\text{hypothesis }0:~~ (\Xm Y) \sim P_{\Xm Y}\\
&\text{hypothesis }1:~~(\Xm Y) \sim Q_{\Xm} \times Q_Y.
\end{align}
The binary test $T$ is chooses hypothesis $0$ as
\beq
T(0|x_1,\dots,x_L,y) = P_{\hat\Xm |Y}(x_1,\dots,x_L|y)
\eeq
and chooses hypothesis $1$ in all other cases. Thus, the pairwise error probabilities are given by
\begin{align}
	\epsilon_0(T,P) &= 1 - \PP \big[\hat{H} = 0|H=0\big] \\
				& = 1 - \sum_{(x_1,\dotsc,x_L)\in\Xc^L, y\in\Yc} P_{\Xm Y} (x_1,\dotsc,x_L,y) T(0|x_1,\dots,x_L,y)\\
				& = 1 - \sum_{(x_1,\dotsc,x_L)\in\Xc^L, y\in\Yc} P_{\Xm Y} (x_1,\dotsc,x_L,y) P_{\hat\Xm |Y}(x_1,\dots,x_L|y) \label{eq:type1err}\\
				&= 1- \frac{1}{\binom{M-1}{L-1}}(1-\bar\epsilon)\label{eq:type1err2}
\end{align}
and
\begin{align}
	\epsilon_1(T,Q) &= \QQ \big[\hat{H} = 0|H=1\big] \\
	&=  \sum_{(x_1,\dotsc,x_L)\in\Xc^L, y\in\Yc}  Q_{\Xm}(x_1,\dotsc,x_L) Q_Y(y)T(0|x_1,\dots,x_L,y)\\
	&= \sum_{(x_1,\dotsc,x_L)\in\Xc^L, y\in\Yc}  Q_{\Xm}(x_1,\dotsc,x_L) Q_Y(y)P_{\hat\Xm |Y}(x_1,\dots,x_L|y)\label{eq:type2err1}\\
	&= \sum_{(x_1,\dotsc,x_L)\in\Xc^L, y\in\Yc} \frac{1}{\binom M L} Q_Y(y)P_{\hat\Xm |Y}(x_1,\dots,x_L|y)\label{eq:type2err2}\\
	&= \sum_{y\in\Yc} \frac{1}{\binom M L} Q_Y(y)  \sum_{(x_1,\dotsc,x_L)\in\Xc^L} P_{\hat\Xm |Y}(x_1,\dots,x_L|y)\\
	& = \sum_{y\in\Yc} \frac{1}{\binom M L} Q_Y(y)\\
	& = \frac{1}{\binom M L}
\end{align}
where \eqref{eq:type1err} and \eqref{eq:type2err1} follow from the definition of the binary test $T(0|x_1,\dots,x_L,y)$ and \eqref{eq:type2err2} from the definition of $Q_\Xm$ in \eqref{eq:qdist}.

Therefore, from the conditions above we can see that for any distribution $Q_Y$ we have
\begin{align}
\frac{1}{\binom{M-1}{L-1}}(1 - \bar \epsilon) \leq 1 - \alpha_{\frac{1}{\binom{M}{L} }}(P_{\Xm Y} , Q_{\Xm} \times Q_Y) 
\end{align}
since the pairwise error probability $\epsilon_0(T,P)$ of the above binary test cannot be better than the Neyman-Pearson optimal tradeoff \eqref{Neyman-Pearson}.
This proves \eqref{eq:mcbound}. In addition, since $Q_Y$ is arbitrary, this also holds for the maximizing distribution,
\begin{align}
	\frac{1}{\binom{M-1}{L-1}}(1 - \bar \epsilon) \leq 1 - \max_{Q_Y} \alpha_{\frac{1}{\binom{M}{L} }}(P_{\Xm Y} , Q_{\Xm} \times Q_Y).
\end{align}

In order to prove the tightness of the bound, we now need to show  that
\begin{align} \label{crgbeifao}
	\frac{1}{\binom{M-1}{L-1}}(1 - \bar \epsilon) 
	&\geq 1 - \max_{Q_Y} \alpha_{\frac{1}{\binom{M}{L} }}(P_{\Xm Y} , Q_{\Xm} \times Q_Y).
\end{align}
In order to show \eqref{crgbeifao} we set $Q_Y=Q_Y^*$ defined in \eqref{eq:qstar} and rewrite the $\alpha_\beta$ function as (see e.g. \cite[Ch.  11]{verdu_IT_book})
\begin{align}
	\alpha_{\frac{1}{\binom{M}{L}}}(P_{\Xm Y} , Q_{\Xm} \times Q^*_Y) =  \sup_{\lambda \geq 0} \left\{ \PP \left[ \frac{P_{\Xm Y}(\Xm,Y)}{Q_{\Xm} (\Xm)\times Q^*_Y(Y)} \leq \lambda \right] + \lambda\QQ \left[ \frac{P_{\Xm Y}(\Xm,Y)}{Q_{\Xm} (\Xm)\times Q^*_Y(Y)} > \lambda \right] - \frac{1}{\binom{M}{L}}\lambda \right\}
\end{align}
where the first probability is computed with respect to $P_{\Xm Y}$ and the second one is computed with respect to $Q_{\Xm} \times Q_Y$. If we  set 
\beq
\lambda^* = \binom{M}{L} \sum_{y \in \Yc} \max_{(x_1,\dotsc,x_L) \in \Xc^L} P_{\Xm Y}(x_1,\dotsc,x_L,y) 
\label{eq:lambdas}
\eeq
we find that
\begin{align}
	  \PP \left[ \frac{P_{\Xm Y}(\Xm,Y)}{Q_{\Xm} (\Xm)\times Q^*_Y(Y)} \leq \lambda^* \right] + \lambda^*\QQ \left[ \frac{P_{\Xm Y}(\Xm,Y)}{Q_{\Xm} (\Xm)\times Q^*_Y(Y)} > \lambda^* \right] - \frac{1}{\binom{M}{L}}\lambda^* 
	 & = 1 - \frac{1}{\binom{M}{L}}\lambda^* \label{eq:vhstep1} 
\end{align}
where \eqref{eq:vhstep1} follows since
\begin{align}
 &Q_{\Xm} (\Xm) Q^*_Y(Y) \lambda^* \\
 & = \frac{1}{\binom{M}{L} }  \frac{\max_{( x_1,\dotsc, x_L) \in \Xc^L} P_{ \Xm Y}( x_1,\dotsc, x_L,y)}{\sum_{y \in \Yc} \max_{( x_1,\dotsc, x_L) \in \Xc^L} P_{ \Xm Y}( x_1,\dotsc, x_L,y)} \binom{M}{L} \sum_{y \in \Yc} \max_{(x_1,\dotsc,x_L) \in \Xc^L} P_{\Xm Y}(x_1,\dotsc,x_L,y)\\
 &= \max_{( x_1,\dotsc, x_L) \in \Xc^L} P_{ \Xm Y}( x_1,\dotsc, x_L,y)
\end{align}
implying that
\begin{align}
\PP \left[ \frac{P_{\Xm Y}(\Xm,Y)}{Q_{\Xm} (\Xm)\times Q^*_Y(Y)} \leq \lambda^* \right] &= \PP \left[P_{\Xm Y}(\Xm,Y) \leq  \max_{( x_1,\dotsc, x_L) \in \Xc^L} P_{ \Xm Y}( x_1,\dotsc, x_L,y) \right] =1\label{eq:pinfspec1}\\
\QQ \left[ \frac{P_{\Xm Y}(\Xm,Y)}{Q_{\Xm} (\Xm)\times Q^*_Y(Y)} > \lambda^* \right] &= \QQ \left[P_{\Xm Y}(\Xm,Y) >  \max_{( x_1,\dotsc, x_L) \in \Xc^L} P_{ \Xm Y}( x_1,\dotsc, x_L,y) \right] =0.\label{eq:pinfspec2}
\end{align}

From Lemma \ref{lemma:minpe}, we have that
\beq
\bar\epsilon=1-\binom{M-1}{L-1}\sum_{y \in \Yc} \max_{(\hat x_1,\dotsc,\hat x_L) \in \Xc^L} P_{ \Xm Y}(\hat x_1,\dotsc,\hat x_L,y)
\eeq
and thus,
\begin{align}
1-\frac{1}{\binom{M}{L}}\lambda^* &=  1 -  \sum_{y \in \Yc} \max_{(x_1,\dotsc,x_L) \in \Xc^L} P_{\Xm Y}(x_1,\dotsc,x_L,y)\\
&=1 - \frac{1}{\binom{M-1}{L-1}} (1 - \bar \epsilon),
\end{align}
which implies that
\begin{align}
	\alpha_{\frac{1}{\binom{M}{L}}}(P_{\Xm Y} , Q_{\Xm} \times Q^*_Y) &=  \sup_{\lambda \geq 0} \left\{ \PP \left[ \frac{P_{\Xm Y}(\Xm,Y)}{Q_{\Xm} (\Xm)\times Q^*_Y(Y)} \leq \lambda \right] + \lambda\QQ \left[ \frac{P_{\Xm Y}(\Xm,Y)}{Q_{\Xm} (\Xm)\times Q^*_Y(Y)} > \lambda \right] - \frac{1}{\binom{M}{L}}\lambda \right\}\label{eq:alphabetalong1}\\
	& \geq \PP \left[ \frac{P_{\Xm Y}(\Xm,Y)}{Q_{\Xm} (\Xm)\times Q^*_Y(Y)} \leq \frac{1}{\binom{M}{L}}\lambda^* \right] + \lambda^*\QQ \left[ \frac{P_{\Xm Y}(\Xm,Y)}{Q_{\Xm}(\Xm) \times Q^*_Y(Y)} > \lambda^* \right] - \lambda^*\label{eq:alphabetalong2} \\
	&= 1 - \frac{1}{\binom{M-1}{L-1}} (1 - \bar \epsilon)\label{eq:alphabetalong3}
\end{align}
proving the desired result. 

\end{IEEEproof}

The identity established by Theorem \ref{thm:exact} can be rewritten in terms of the alternative pairwise error probability tradeoff.

\begin{corollary}
Identity \eqref{eq:mc} can be rewritten as
\beq
\frac{1}{\binom{M}{L}} = \max_{Q_Y} \beta_{1 - \frac{1}{\binom{M-1}{L-1}} (1 - \bar \epsilon)}(P_{\Xm Y} , Q_{\Xm} \times Q_Y).
\eeq
\end{corollary}

The proof of Theorem \ref{thm:exact} suggests a broad family of lower bounds to the probability of error parametrized by the auxiliary distribution $Q_Y$. In particular, we have that for a fixed auxiliary distribution $Q_Y$, we have that
\begin{align}
\frac{1}{\binom{M-1}{L-1}}(1 - \bar \epsilon) &\leq 1 - \alpha_{\frac{1}{\binom{M}{L} }}(P_{\Xm Y} , Q_{\Xm} \times Q_Y),
\end{align}
or equivalently,
\begin{align}
\frac{1}{\binom{M}{L}} &\geq \beta_{1 - \frac{1}{\binom{M-1}{L-1}} (1 - \bar \epsilon)}(P_{\Xm Y} , Q_{\Xm} \times Q_Y).
\end{align}

In order to efficiently compute these bounds, one must choose a convenient $Q_Y$. The specific choice will, naturally, depend on the specifics of the problem at hand.

\section{Information Spectrum}
\label{sec:vh}

In this section, we show an alternative identity for the probability of error of list hypothesis testing. Specifically, this identity is expressed as a function of the tail probability that the likelihood ratio exceeds a certain threshold. These expressions have sometimes been termed information spectrum \cite{han2003information}.

\begin{theorem}
For a fixed auxiliary distribution $Q_Y$ and constant $\lambda\geq0$, we have that
\begin{align}
\frac{1}{\binom{M-1}{L-1}}(1 - \bar \epsilon) &\leq 1- \left\{ \PP \left[ \frac{P_{\Xm Y}(\Xm,Y)}{Q_{\Xm}(\Xm) \times Q_Y(Y)} \leq \lambda \right]  - \frac{1}{\binom{M}{L}}\lambda \right\}.\label{eq:vhbound}
\end{align}
In addition, 
\begin{align}
\frac{1}{\binom{M-1}{L-1}}(1 - \bar \epsilon) &= 1-\max_{Q_Y}\sup_{\lambda \geq 0 } \left\{ \PP \left[ \frac{P_{\Xm Y}(\Xm,Y)}{Q_{\Xm}(\Xm) \times Q_Y(Y)} \leq \lambda \right]  - \frac{1}{\binom{M}{L}}\lambda \right\}\label{eq:vh_qopt}\\
&= 1-\sup_{\lambda \geq 0 } \left\{ \PP \left[ \frac{P_{\Xm Y}(\Xm,Y)}{Q_{\Xm}(\Xm) \times Q^*_Y(Y)} \leq \lambda \right]  - \frac{1}{\binom{M}{L}}\lambda \right\}
\end{align}
where $Q^*_Y$ defined in \eqref{eq:qstar} is the unique maximizer of  \eqref{eq:vh_qopt}.
\end{theorem}

\begin{IEEEproof}
As shown in the proof of Theorem \ref{thm:exact}, we have that, 
\begin{align}
\alpha_{\frac{1}{\binom{M}{L}}}(P_{\Xm Y} , Q_{\Xm} \times Q_Y) &= \sup_{\lambda \geq 0} \left\{ \PP \left[ \frac{P_{\Xm Y}(\Xm,Y)}{Q_{\Xm} (\Xm)\times Q_Y(Y)} \leq \lambda \right] + \lambda\QQ \left[ \frac{P_{\Xm Y}(\Xm,Y)}{Q_{\Xm} (\Xm)\times Q_Y(Y)} > \lambda \right] - \frac{1}{\binom{M}{L}}\lambda \right\}\label{eq:longalphavh22}\\
&\geq \PP \left[ \frac{P_{\Xm Y}(\Xm,Y)}{Q_{\Xm} (\Xm)\times Q_Y(Y)} \leq \lambda \right] + \lambda\QQ \left[ \frac{P_{\Xm Y}(\Xm,Y)}{Q_{\Xm} (\Xm)\times Q_Y(Y)} > \lambda \right] - \frac{1}{\binom{M}{L}}\lambda\label{eq:longalphavh}\\
&\geq \PP \left[ \frac{P_{\Xm Y}(\Xm,Y)}{Q_{\Xm} (\Xm)\times Q_Y(Y)} \leq \lambda \right] - \frac{1}{\binom{M}{L}}\lambda\label{eq:longalphavh2}
\end{align}
where \eqref{eq:longalphavh} holds for any fixed $\lambda\geq0$ and \eqref{eq:longalphavh2} follows since the second term is always non-negative. Applying this to \eqref{eq:mcbound}, the bound \eqref{eq:vhbound} follows.

For the particular choice $\lambda^*$ in \eqref{eq:lambdas}
\begin{align}
\frac{1}{\binom{M-1}{L-1}} (1 - \bar \epsilon)&= 1- \sup_{\lambda \geq 0} \left\{ \PP \left[ \frac{P_{\Xm Y}(\Xm,Y)}{Q_{\Xm} (\Xm)\times Q^*_Y(Y)} \leq \lambda \right] + \lambda\QQ \left[ \frac{P_{\Xm Y}(\Xm,Y)}{Q_{\Xm} (\Xm)\times Q^*_Y(Y)} > \lambda \right] - \frac{1}{\binom{M}{L}}\lambda \right\}\label{eq:vhlambdas1}\\
&= 1 -  \PP \left[ \frac{P_{\Xm Y}(\Xm,Y)}{Q_{\Xm}(\Xm) \times Q^*_Y(Y)} \leq \lambda^* \right]  - \frac{1}{\binom{M}{L}}\lambda^* \label{eq:vhlambdas}\\
&= 1 - 1 + \frac{1}{\binom{M-1}{L-1}} (1 - \bar \epsilon) \label{eq:vhlambdas2}\\
&=\frac{1}{\binom{M-1}{L-1}} (1 - \bar \epsilon)\label{eq:vhlambdas3}
\end{align}
where \eqref{eq:vhlambdas} and \eqref{eq:vhlambdas2} follow from \eqref{eq:pinfspec1} and \eqref{eq:pinfspec2}, respectively. Eqs. \eqref{eq:vhlambdas1}-\eqref{eq:vhlambdas3} imply that $\lambda^*$ in \eqref{eq:lambdas} is a maximizer of \eqref{eq:vhlambdas1}, and thus
\begin{align}
	\frac{1}{\binom{M-1}{L-1}} (1 - \bar \epsilon) 
	&=  1 - \sup_{\lambda \geq 0 } \left\{ \PP \left[ \frac{P_{\Xm Y}(\Xm,Y)}{Q_{\Xm}(\Xm) \times Q^*_Y(Y)} \leq \lambda \right]  - \frac{1}{\binom{M}{L}}\lambda \right\}.
\end{align}

We now proceed with the proof that $Q^*_Y$ as defined in \eqref{eq:qstar} is the unique maximizer of \eqref{eq:vh_qopt}. We divide the proof in two parts, depending on whether or not $Q_\Xm \times Q^*_Y$ is absolutely continuous with respect to $P_{\Xm Y}$.

\vspace{2mm}
\ul{$Q_\Xm \times Q^*_Y$ is absolutely continuous with respect to $P_{\Xm Y}$}

Using \eqref{eq:longalphavh22}, we rewrite \eqref{eq:mc} as
\beq
\frac{1}{\binom{M-1}{L-1}} (1 - \bar \epsilon) = 1-\max_{Q_Y}\sup_{\lambda \geq 0} \left\{ \PP \left[ \frac{P_{\Xm Y}(\Xm,Y)}{Q_{\Xm} (\Xm)\times Q_Y(Y)} \leq \lambda \right] + \lambda\QQ \left[ \frac{P_{\Xm Y}(\Xm,Y)}{Q_{\Xm} (\Xm)\times Q_Y(Y)} > \lambda \right] - \frac{1}{\binom{M}{L}}\lambda \right\}.\label{eq:vhlonglong}
\eeq
The above expression and \eqref{eq:vh_qopt} are both exact characterizations of the error probability. However, \eqref{eq:vhlonglong} has an additional non-negative term compared to \eqref{eq:vh_qopt}. Thus, any maximizing distribution and constant $Q_Y^*$ and $\lambda^*$ of \eqref{eq:vh_qopt} are also mazimizers of \eqref{eq:vhlonglong}. As a result, by comparing both equations, we have that
\beq
\QQ \left[ \frac{P_{\Xm Y}(\Xm,Y)}{Q_{\Xm} (\Xm)\times Q^*_Y(Y)} > \lambda^* \right] =0
\eeq
Using the definition of $Q_{\Xm}$ in \eqref{eq:qdist}, and the absolute continuity of $Q_\Xm \times Q^*_Y$ with respect to $P_{\Xm Y}$ this implies that 
\beq
P_{\Xm Y}(x_1,\dotsc,x_L,y) \leq \frac{\lambda^*}{\binom M L}\,Q^*_Y(y)
\eeq
for all $(x_1,\dotsc,x_L)\in\Xc^L$ and $y\in\Yc$.  Since this expression holds for arbitrary $(x_1,\dotsc,x_L)\in\Xc^L$, in particular it holds for the maximizing $(x_1,\dotsc,x_L)\in\Xc^L$, yielding
\beq
\max_{(x_1,\dotsc,x_L)\in\Xc^L}P_{\Xm Y}(x_1,\dotsc,x_L,y) \leq \frac{\lambda^*}{\binom M L} \,Q^*_Y(y).
\label{eq:vhineqq}
\eeq
Summing over $y\in\Yc$ yields
\begin{align}
\sum_{y\in\Yc}\max_{(x_1,\dotsc,x_L)\in\Xc^L}P_{\Xm Y}(x_1,\dotsc,x_L,y) &\leq \frac{\lambda^*}{\binom M L}\sum_{y\in\Yc}Q^*_Y(y)\\
&=\frac{\lambda^*}{\binom M L}
\label{eq:sumqvh}
\end{align}
where \eqref{eq:sumqvh} follows from the fact that $Q_Y^*$ is a probability distribution.

We have shown that for the maximizing $Q_Y^*$, $\lambda^*$ must satisfy \eqref{eq:sumqvh}. In addition, the maximizing $\lambda^*$ must minimize the second term of \eqref{eq:vh_qopt}.

Therefore, since the first term of \eqref{eq:vh_qopt} is increasing with $\lambda$, for any $\lambda$ satisfying \eqref{eq:sumqvh}, the smallest $\lambda$ satisfying \eqref{eq:sumqvh} is the maximizer of \eqref{eq:vh_qopt}, and thus, equality in \eqref{eq:sumqvh} must hold, i.e.,
\begin{align}
\sum_{y\in\Yc}\max_{(x_1,\dotsc,x_L)\in\Xc^L}P_{\Xm Y}(x_1,\dotsc,x_L,y) &=\frac{\lambda^*}{\binom M L}
\label{eq:sumqvh2}
\end{align}
Substituting $\lambda^*$ in \eqref{eq:sumqvh2} into \eqref{eq:vhineqq} yields
\beq
\frac{\max_{(x_1,\dotsc,x_L)\in\Xc^L}P_{\Xm Y}(x_1,\dotsc,x_L,y)}{\sum_{y\in\Yc}\max_{(x_1,\dotsc,x_L)\in\Xc^L}P_{\Xm Y}(x_1,\dotsc,x_L,y)} \leq Q^*_Y(y).
\label{eq:vhoptqineq}
\eeq
Observe that the left hand side of \eqref{eq:vhoptqineq} is itself a probability distribution on $\Yc$ and thus, \eqref{eq:vhoptqineq} holds with equality for all $y\in\Yc$, recovering \eqref{eq:qstar}.

\vspace{2mm}

\ul{$Q_\Xm \times Q^*_Y$ is not absolutely continuous with respect to $P_{\Xm Y}$}

Consider a distribution $V_Y$ on $\Yc$ and a non-Bayesian binary hypothesis test between $P_{\Xm Y}$ and $Q_\Xm\times V_Y$. Then, if there exists some $\hat y\in\Yc$ such that $V_Y(\hat y) = 0$, any optimal test in the Neyman-Pearson setting $T$ is such that
\beq
T(1|x_1,\dotsc,x_L,\hat y) \cdot P_{\Xm Y}(x_1,\dotsc,x_L,\hat y) = 0
\eeq
for every $(x_1,\dotsc,x_L)\in\Xc^L$. The interpretation of this statement is that whenever $V_Y(\hat y)=0$, any optimal test would not choose hypothesis $1$, unless $P_{\Xm Y}(x_1,\dotsc,x_L,\hat y) = 0$ for all $(x_1,\dotsc,x_L)\in\Xc^L$.

We have the following result, whose proof can be found in Appendix \ref{app:prlemqhat}.

\begin{lemma}
\label{lem:qhat}
Let $Q_Y$ be a distribution on $\Yc$. If there exists a $\bar y$ such that
\begin{enumerate}
\item $Q_Y(\bar y) = 0$
\item $\exists \xv^{1},\xv^{2}\in\Xc^L$, with $\xv^i = (x_1^i,\dotsc,x_L^i)$ such that $P_{\Xm Y}(\xv^1,\bar y) P_{\Xm Y}(\xv^2,\bar y) >0$,
\end{enumerate}
then, there exists a distribution $\hat Q_Y$ on $\Yc$ such that
\beq
\alpha_{\frac{1}{\binom{M}{L}}} (P_{\Xm Y},Q_\Xm \times Q_Y) < \alpha_{\frac{1}{\binom{M}{L}}} (P_{\Xm Y},Q_\Xm \times \hat Q_Y).
\eeq

\end{lemma}

The above Lemma shows that, if there are two (or more) hypotheses for which $P_{\Xm Y}(\xv^1,\bar y) P_{\Xm Y}(\xv^2,\bar y) >0$, an auxiliary distribution $Q_Y$ that associates zero mass to observation $\bar y$ cannot be optimal. In particular, the lemma shows the existence of a distribution that places non-zero mass to all $y\in\Yc$ that is better than one that places zero mass at $\bar y$, thus bringing us back to the case where $Q_\Xm \times Q^*_Y$ is absolutely continuous with respect to $P_{\Xm Y}$.

There is a remaining trivial case, where there are observations $y\in\Yc$ that can only be obtained from only one individual hypothesis. In this case, there is no ambiguity as to what hypothesis caused the observation. Thus, then the problem reduces to removing those observations, i.e., the optimal distribution places zero mass on those and non-zero on the others.

\end{IEEEproof}

\appendices

\section{Proof of Lemma \ref{lem:qhat}}
\label{app:prlemqhat}

Let $(T,\lambda^*)$ be an optimal non-Bayesian likelihood-ratio test and the corresponding threshold for testing between $P_{\Xm Y}$ and $Q_\Xm\times Q_Y$ with fixed type-1 error probability $\epsilon_1(T,Q_\Xm\times Q_Y) = \frac{1}{\binom ML}$.

Consider the distribution $\hat Q_Y$
\beq
\hat Q_Y(y) = \begin{cases}
\frac{\binom ML}{\mu}\max_{( x_1,\dotsc, x_L) \in \Xc^L} P_{ \Xm Y}( x_1,\dotsc, x_L,y) & y = \bar y\\
\frac{\lambda^*}{\mu}Q_Y(y) & y\neq \bar y
\end{cases}
\label{eq:hatQ2}
\eeq
where $\mu = \binom ML\max_{( x_1,\dotsc, x_L) \in \Xc^L} P_{ \Xm Y}( x_1,\dotsc, x_L,y) + \lambda^*$.

We first show that $\hat Q_Y$ is a probability distribution on $\Yc$, i.e., that
\beq
\sum_y \hat Q_Y(y) = 1.
\eeq
We write 
\begin{align}
\sum_y \hat Q_Y(y) &= \sum_{y\neq\bar y} \hat Q_Y(y) + \hat Q_Y(\bar y)\\
&= \frac{\lambda^*}{\mu}\sum_{y\neq\bar y}  Q_Y(y)+  \frac{\binom ML}{\mu}\max_{( x_1,\dotsc, x_L) \in \Xc^L} P_{ \Xm Y}( x_1,\dotsc, x_L,y) \label{eq:defqh}\\
& = \frac1\mu \bigg(\binom ML\cdot \max_{( x_1,\dotsc, x_L) \in \Xc^L} P_{ \Xm Y}( x_1,\dotsc, x_L,y) + \lambda^*\bigg)\label{eq:qsum1}\\
&=1 \label{eq:endsum1},
\end{align}
where \eqref{eq:defqh} follows from the definition of $\hat Q_Y$, \eqref{eq:qsum1} follows from the fact that $Q_Y$ is a probability distribution with $Q_Y(\bar y) = 0$, and \eqref{eq:endsum1} follows from the definition of $\mu$.

We now proceed with the proof that for the distribution $\hat Q_Y$ in \eqref{eq:hatQ2} we have that
\beq
\alpha_{\frac{1}{\binom{M}{L}}} (P_{\Xm Y},Q_\Xm \times Q_Y) < \alpha_{\frac{1}{\binom{M}{L}}} (P_{\Xm Y},Q_\Xm \times \hat Q_Y).
\eeq
In particular, we construct a binary test $\hat T$ to test between $P_{\Xm Y}$ and $Q_\Xm\times \hat Q_Y$ and show that
\begin{align}
&\epsilon_1(\hat T,Q_\Xm\times \hat Q_Y) = \frac{1}{\binom ML} \label{eq:epseq1}\\
&\epsilon_0(\hat T,P_{\Xm Y}) >\epsilon_0( T,P_{\Xm Y}).\label{eq:epseq2}
\end{align}
In addition, if we show that the test $\hat T$ is an optimal test in the Neyman-Pearson sense, the proof will be complete.

Consider the set defined in \eqref{eq:setS}
\beq
\Sc(y) = \Big\{( x_1,\dotsc, x_L)\in\Xc^L \ : \  P_{ \Xm Y} ( x_1,\dotsc, x_L,y) = \max_{(\hat x_1,\dotsc,\hat x_L)\in\Xc^L}P_{ \Xm Y}(\hat x_1,\dotsc,\hat x_L,y) \Big\}
\eeq
and let 
\beq
\bar \xv=(\bar x_1,\dotsc,\bar x_L)\in\Sc(y).
\label{eq:defxbar}
\eeq 
We construct the test $\hat T$ as follows
\beq
\hat T(1|x_1,\dotsc,x_L,y) = \begin{cases}
T(1|x_1,\dotsc,x_L,y) & y\neq \bar y \\
1 & y=\bar y, ( x_1,\dotsc, x_L)\neq (\bar x_1,\dotsc,\bar x_L) \\
0 & y=\bar y, ( x_1,\dotsc, x_L)= (\bar x_1,\dotsc,\bar x_L).
\end{cases}
\label{eq:testhatdef}
\eeq

We next validate equality \eqref{eq:epseq1},
\begin{align}
1- &\epsilon_1(\hat T,Q_\Xm\times \hat Q_Y)\notag\\
 & = \sum_{y\in\Yc}\sum_{(x_1,\dotsc,x_L)\in\Xc^L} Q_\Xm(x_1,\dotsc,x_L) \hat Q_Y(y) \hat T(1|x_1,\dotsc,x_L,y)\\
& =\sum_{y\neq \bar y}\sum_{(x_1,\dotsc,x_L)\in\Xc^L} Q_\Xm(x_1,\dotsc,x_L) \hat Q_Y(y) \hat T(1|x_1,\dotsc,x_L,y) \notag\\
&~~~~~~+ \sum_{(x_1,\dotsc,x_L)\in\Xc^L} Q_\Xm(x_1,\dotsc,x_L) \hat Q_Y(\bar y) \hat T(1|x_1,\dotsc,x_L,\bar y)\\
& =\sum_{y\neq \bar y}\sum_{(x_1,\dotsc,x_L)\in\Xc^L} Q_\Xm(x_1,\dotsc,x_L) \frac{\lambda^*}{\mu}Q_Y(y) \hat T(1|x_1,\dotsc,x_L,y) \notag\\
&~~+ \sum_{\substack{(x_1,\dotsc,x_L)\in\Xc^L \\ (x_1,\dotsc,x_L)\neq (\bar x_1,\dotsc,\bar x_L)}} Q_\Xm(x_1,\dotsc,x_L) \frac{\binom ML}{\mu}\max_{( x_1,\dotsc, x_L) \in \Xc^L} P_{ \Xm Y}( x_1,\dotsc, x_L,\bar y) \hat T(1|x_1,\dotsc,x_L,\bar y) \label{eq:eps1long1}\\
& =\frac{\lambda^*}{\mu} \sum_{y\neq \bar y}\sum_{(x_1,\dotsc,x_L)\in\Xc^L} Q_\Xm(x_1,\dotsc,x_L) Q_Y(y)  T(1|x_1,\dotsc,x_L,y) \notag\\
&~~~~~~+\frac{\binom ML}{\mu}\max_{( x_1,\dotsc, x_L) \in \Xc^L} P_{ \Xm Y}( x_1,\dotsc, x_L,y) \sum_{\substack{(x_1,\dotsc,x_L)\in\Xc^L \\ (x_1,\dotsc,x_L)\neq (\bar x_1,\dotsc,\bar x_L)}} Q_\Xm(x_1,\dotsc,x_L) \label{eq:eps1long2}\\
&= \frac{\lambda^*}{\mu} \big(1- \epsilon_1(T,Q_\Xm\times \hat Q_Y)\big) + \frac{\binom ML}{\mu}\max_{( x_1,\dotsc, x_L) \in \Xc^L} P_{ \Xm Y}( x_1,\dotsc, x_L,y) \bigg(1 - \frac{1}{\binom ML}\bigg) \label{eq:eps1long3}\\
&= \frac{\lambda^*}{\mu} \bigg(1 - \frac{1}{\binom ML}\bigg) + \frac{\binom ML}{\mu}\max_{( x_1,\dotsc, x_L) \in \Xc^L} P_{ \Xm Y}( x_1,\dotsc, x_L,y) \bigg(1 - \frac{1}{\binom ML}\bigg)\label{eq:eps1long4}\\
&= 1 - \frac{1}{\binom ML}.\label{eq:eps1long5}
\end{align}
where \eqref{eq:eps1long1} follows from the definition of the distribution $\hat Q_Y$ in \eqref{eq:hatQ2}, \eqref{eq:eps1long2} from the definition of the test $\hat T$ in \eqref{eq:testhatdef}, \eqref{eq:eps1long3} follows from the definition of the type-1 probability of error for test $T$ and from the definition of the distribution $Q_\Xm$ in \eqref{eq:qdist}, \eqref{eq:eps1long4} follows from the the fact that by the definition of test $T$, its type-1 error probability is $\frac{1}{\binom ML}$, and \eqref{eq:eps1long5} follows from the definition of the normalization constant $\mu$. 

Now, we turn to inequality \eqref{eq:epseq2}. We have that
\begin{align}
\epsilon_0(T,P_{\Xm Y}) &=  \sum_{y\in\Yc}\sum_{(x_1,\dotsc,x_L)\in\Xc^L} P_{\Xm Y} (x_1,\dotsc,x_L,y) T(1|x_1,\dotsc,x_L,y)\\
&=  \sum_{y\neq \bar y}\sum_{(x_1,\dotsc,x_L)\in\Xc^L} P_{\Xm Y} (x_1,\dotsc,x_L,y) T(1|x_1,\dotsc,x_L,y) \notag \\
&~~~~~~+  \sum_{(x_1,\dotsc,x_L)\in\Xc^L} P_{\Xm Y} (x_1,\dotsc,x_L,\bar y) T(1|x_1,\dotsc,x_L,\bar y)\label{eq:eps0long1}\\
&< \sum_{y\neq \bar y}\sum_{(x_1,\dotsc,x_L)\in\Xc^L} P_{\Xm Y} (x_1,\dotsc,x_L,y) T(1|x_1,\dotsc,x_L,y) \notag \\
&~~~~~~+  \sum_{\substack{(x_1,\dotsc,x_L)\in\Xc^L\\(x_1,\dotsc,x_L)\neq (\bar x_1,\dotsc,\bar x_L)}} P_{\Xm Y} (x_1,\dotsc,x_L,\bar y) \label{eq:eps0long2}\\
&= \sum_{y\neq \bar y}\sum_{(x_1,\dotsc,x_L)\in\Xc^L} P_{\Xm Y} (x_1,\dotsc,x_L,y) \hat T(1|x_1,\dotsc,x_L,y) \notag \\
&~~~~~~+  \sum_{\substack{(x_1,\dotsc,x_L)\in\Xc^L\\(x_1,\dotsc,x_L)\neq (\bar x_1,\dotsc,\bar x_L)}} P_{\Xm Y} (x_1,\dotsc,x_L,\bar y)\hat T(1|x_1,\dotsc,x_L,\bar y)\label{eq:eps0long3}\\
&=\epsilon_0(\hat T,P_{\Xm Y})\label{eq:eps0long4}
\end{align}
where \eqref{eq:eps0long1} follows from splitting the sum over $y$ in two, \eqref{eq:eps0long2} follows from the fact that for $\bar y$ the test $T$ is such that $T(1|x_1,\dotsc,x_L,\bar y) P_{\Xm Y} (x_1,\dotsc,x_L,\bar y) =0$, and from the fact that there are at least two hypotheses for which $P_{\Xm Y} (x_1,\dotsc,x_L,\bar y)\neq 0$ from the statement of the lemma; we thus upper bound it by a non-zero term. Finally, \eqref{eq:eps0long2} follows from the definition of $\hat T$ in \eqref{eq:testhatdef} and \eqref{eq:eps0long4} follows from the definition of type-0 probability of error.

We now will be done if we show that the test $\hat T$ is an optimal test in the Neyman-Pearson sense. Since we have shown that $\epsilon_1(\hat T,Q_\Xm\times \hat Q_Y) = \frac{1}{\binom ML}$, all we need to show is that there exists a threshold for the likelihood ratio. We will now show that $\mu$ is indeed this threshold for test $\hat T$. We divide the proof in several cases
\begin{itemize}
\item When $y\neq \bar y$, we have that
\begin{align}
\frac{P_{\Xm Y} (x_1,\dotsc,x_L,y)}{\mu Q_\Xm(x_1,\dotsc,x_L) \hat Q_Y(y)} &= \frac{P_{\Xm Y} (x_1,\dotsc,x_L,y)}{\mu Q_\Xm(x_1,\dotsc,x_L) \frac{\lambda^*}{\mu} Q_Y(y)}\label{eq:testhatq}\\
&= \frac{P_{\Xm Y} (x_1,\dotsc,x_L,y)}{\lambda^* Q_\Xm(x_1,\dotsc,x_L) Q_Y(y)}, 
\end{align}
where \eqref{eq:testhatq} follows from the definition of the distribution $\hat Q_Y$ in \eqref{eq:hatQ2}. Thus, since $\lambda^*$ was an optimal threshold for test $T$, $\mu$ is an optimal threshold for test $\hat T$ in this case.
\item When $y= \bar y, (x_1,\dotsc,x_L)\neq (\bar x_1,\dotsc,\bar x_L)$ according to the definition of $\hat T$ in \eqref{eq:testhatdef}, we have that $ \hat T(1| x_1,\dotsc, x_L,\bar y) = 1$. In addition,
\begin{align}
\mu Q_\Xm( x_1,\dotsc, x_L) \hat Q(\bar y) &= \mu Q_\Xm( x_1,\dotsc, x_L) \frac{\binom ML}{\mu}\max_{( x_1,\dotsc, x_L) \in \Xc^L} P_{ \Xm Y}( x_1,\dotsc, x_L,\bar y)\label{eq:testhatq1}\\
&=\max_{( x_1,\dotsc, x_L) \in \Xc^L} P_{ \Xm Y}( x_1,\dotsc, x_L,\bar y)\label{eq:testhatq11}\\
&\geq P_{ \Xm Y}( x_1,\dotsc, x_L,\bar y)
\end{align}
where \eqref{eq:testhatq1} follows from the definition of $\hat Q_Y$ in \eqref{eq:hatQ2} and \eqref{eq:testhatq11} from the definitions of $Q_\Xm$ in \eqref{eq:qdist} and of $(\bar x_1,\dotsc, \bar x_L)$ in \eqref{eq:defxbar}. This implies that when $y= \bar y, (x_1,\dotsc,x_L)\neq (\bar x_1,\dotsc,\bar x_L)$,
\beq
\frac{P_{ \Xm Y}( x_1,\dotsc, x_L,\bar y)}{Q_\Xm( x_1,\dotsc, x_L) \hat Q(\bar y)} \leq \mu.
\eeq

\item When $y= \bar y,(x_1,\dotsc,x_L)= (\bar x_1,\dotsc,\bar x_L)$, according to the definition of $\hat T$ in \eqref{eq:testhatdef}, we have that $ \hat T(1|\bar x_1,\dotsc,\bar x_L,\bar y) = 0$. In addition,
\begin{align}
\mu Q_\Xm(\bar x_1,\dotsc,\bar x_L) \hat Q(\bar y) &= \mu Q_\Xm(\bar x_1,\dotsc,\bar x_L) \frac{\binom ML}{\mu}\max_{( x_1,\dotsc, x_L) \in \Xc^L} P_{ \Xm Y}( x_1,\dotsc, x_L,\bar y)\label{eq:testhatq2}\\
&=P_{ \Xm Y}(\bar x_1,\dotsc, \bar x_L,\bar y)\label{eq:testhatq3}
\end{align}
where \eqref{eq:testhatq2} follows from the definition of $\hat Q_Y$ in \eqref{eq:hatQ2} and \eqref{eq:testhatq3} from the definitions of $Q_\Xm$ in \eqref{eq:qdist} and of $(\bar x_1,\dotsc, \bar x_L)$ in \eqref{eq:defxbar}. This implies that for $y= \bar y,(x_1,\dotsc,x_L)= (\bar x_1,\dotsc,\bar x_L)$, we have that
\beq
\frac{P_{ \Xm Y}(\bar x_1,\dotsc, \bar x_L,\bar y)}{Q_\Xm(\bar x_1,\dotsc,\bar x_L) \hat Q(\bar y)} = \mu.
\eeq
\end{itemize}
As a result, the test $\hat T$ is an optimal Neyman-Pearson test and satisfies
\beq
\alpha_{\frac{1}{\binom{M}{L}}} (P_{\Xm Y},Q_\Xm \times Q_Y) < \alpha_{\frac{1}{\binom{M}{L}}} (P_{\Xm Y},Q_\Xm \times \hat Q_Y).
\eeq


%
%
%
%
%
%


\begin{thebibliography}{1}
\providecommand{\url}[1]{#1}
\csname url@samestyle\endcsname
\providecommand{\newblock}{\relax}
\providecommand{\bibinfo}[2]{#2}
\providecommand{\BIBentrySTDinterwordspacing}{\spaceskip=0pt\relax}
\providecommand{\BIBentryALTinterwordstretchfactor}{4}
\providecommand{\BIBentryALTinterwordspacing}{\spaceskip=\fontdimen2\font plus
\BIBentryALTinterwordstretchfactor\fontdimen3\font minus
  \fontdimen4\font\relax}
\providecommand{\BIBforeignlanguage}[2]{{%
\expandafter\ifx\csname l@#1\endcsname\relax
\typeout{** WARNING: IEEEtran.bst: No hyphenation pattern has been}%
\typeout{** loaded for the language `#1'. Using the pattern for}%
\typeout{** the default language instead.}%
\else
\language=\csname l@#1\endcsname
\fi
#2}}
\providecommand{\BIBdecl}{\relax}
\BIBdecl

\bibitem{lehmann2005testing}
E.~L. Lehmann and J.~P. Romano, \emph{Testing statistical hypotheses}.\hskip
  1em plus 0.5em minus 0.4em\relax Springer, 2005.

\bibitem{neyman1933ix}
J.~Neyman and E.~S. Pearson, ``On the problem of the most efficient tests of
  statistical hypotheses,'' \emph{Philosophical Transactions of the Royal
  Society of London. Series A, Containing Papers of a Mathematical or Physical
  Character}, vol. 231, no. 694-706, pp. 289--337, 1933.

\bibitem{scarlett2021introductory}
J.~Scarlett and V.~Cevher, ``An introductory guide to {Fano}'s inequality with
  applications in statistical estimation,'' in \emph{Information-Theoretic
  Methods in Data Science}, M.~R.~D. Rodrigues and Y.~C. Eldar, Eds.\hskip 1em
  plus 0.5em minus 0.4em\relax Cambridge University Press, 2021.

\bibitem{elias1957list}
P.~Elias, ``List decoding for noisy channels,'' in \emph{Wescon Convention
  Record, Part 2.}\hskip 1em plus 0.5em minus 0.4em\relax IRE, 1957, pp.
  95--104.

\bibitem{studer2010soft}
C.~Studer and H.~B{\"o}lcskei, ``Soft--input soft--output single tree-search
  sphere decoding,'' \emph{IEEE Trans. Inf. Theory}, vol.~56, no.~10, pp.
  4827--4842, 2010.

\bibitem{polyanskiy2010channel}
Y.~Polyanskiy, H.~V. Poor, and S.~Verd\'u, ``Channel coding rate in the finite
  blocklength regime,'' \emph{IEEE Trans. Inf. Theory}, vol.~56, no.~5, pp.
  2307--2359, 2010.

\bibitem{han2003information}
T.~S. Han, \emph{Information-spectrum methods in information theory}.\hskip 1em
  plus 0.5em minus 0.4em\relax Springer, 2003.

\bibitem{vazquez2016bayesian}
G.~Vazquez-Vilar, A.~{Tauste Campo}, A.~{Guill\'{e}n i F\`{a}bregas}, and
  A.~Martinez, ``Bayesian {$M$}-ary hypothesis testing: The meta-converse and
  {Verd\'u-Han} bounds are tight,'' \emph{IEEE Trans. Inf. Theory}, vol.~62,
  no.~5, pp. 2324--2333, 2016.

\bibitem{verdu_IT_book}
S.~Verd\'u, \emph{Information Theory}, (draft) 2021.

\end{thebibliography}
\end{document}